
\def\ie{{\it i.e.$\,\,$}}			
\def\eg{{\it e.g.$\,\,$}}			
\def\im{{\rm{i}}}				
\def\e{{\rm{e\,}}}				

\def\M{{\cal M}}				
\def\U{{\cal U}}				
\def\Zs{{{\bf Z}}}				
\def\Index{{\rm{Index}}}			
\def\dir{{{D\!\!\!\!/}_A}}			
\def\ni{\noindent}
\def\ua{\U_{\alpha}} \def\ub{\U_{\beta}} \def\uc{\U_{\gamma}} 
\def\uab{\U_{\alpha}\cap \U_{\beta}} 
\def\uabc{\U_{\alpha}\cap \U_{\beta}\cap \U_{\gamma}}

\def\aa{A^{\alpha}} \def\ab{A^{\beta}}
\def\pa{\phi^{\alpha}} \def\pb{\phi^{\beta}} \def\pc{\phi^{\gamma}}
\def\cab{\chi^{\alpha\beta}} \def\cba{\chi^{\beta\alpha}} 
\def\cbc{\chi^{\beta\gamma}} \def\cca{\chi^{\gamma\alpha}}

\def\va{V_{\alpha}} \def\vb{V_{\beta}} \def\lab{L_{\alpha\beta}}
\def\lab{L^{\alpha\beta}}\def\lbc{L^{\beta\gamma}}
\def\lca{L^{\gamma\alpha}}
\def\fa{F^{\alpha}}
\def\eabc{\epsilon(\alpha\beta\gamma)}
\magnification=1200
\line{\hfil SWAT/99-233}
\line{\hfil hep-th/9906093}
\vskip 1cm
\centerline{\bf{The Dirac quantisation condition for fluxes on four-manifolds}}
\vskip 1cm
\centerline{Marcos Alvarez \quad and \quad  David I. Olive}
\centerline{Department of Physics, University of Wales Swansea,} 
\centerline{ Singleton Park, Swansea SA2 8PP, U.K.}
\centerline{ {\tt{e-mail: d.olive@swansea.ac.uk, pyma@swansea.ac.uk}}}
\vskip 1cm
\centerline{\bf ABSTRACT}
\medskip A systematic treatment is given of the Dirac quantisation 
condition for electromagnetic fluxes through two-cycles on a four-manifold 
space-time which can be very complicated topologically, provided only
 that it is connected, compact, oriented and smooth. This is sufficient for
the quantised Maxwell theory on it to satisfy electromagnetic duality
properties. The results depend upon whether the complex wave function needed
for the argument is scalar or spinorial in nature. An essential step is the
derivation of a \lq\lq quantum Stokes' theorem" for the integral of the gauge
potential around a closed loop on the manifold. This can only
be done for an exponentiated version of the line integral (the ``Wilson
loop'') and the result again depends  on the nature of the complex wave
functions, through the appearance of what is known as a Stiefel-Whitney
cohomology class in the spinor case. A nice picture emerges providing a
physical interpretation, in terms of quantised fluxes and wave-functions, of
mathematical concepts such as spin structures, spin$_C$ structures, the
Stiefel-Whitney class and Wu's formula. Relations appear between these,
electromagnetic duality and the Atiyah-Singer index theorem. Possible
generalisation to higher dimensions of space-time in the presence of branes is
mentioned.

\bigskip
\ni{\bf 1. Introduction}
\medskip
The main purpose of this paper is to provide a systematic discussion of 
the Dirac quantisation condition for $U(1)$ (electromagnetic) fluxes
through two-cycles on a generic four-manifold $\M_4$. By generic
we mean that $\M_4$ is allowed to possess rather general topological
features subject to the requirements that it be smooth, connected and
Poincar\'e dual. So it is oriented and compact but does not have 
to be simply connected.

Our motivation for doing this is to improve understanding of the 
electromagnetic duality properties of the quantised pure Maxwell 
theory on such manifolds. It does appear that electromagnetic 
duality and Poincar\'e duality of the underlying space-time are
linked. We believe that our results are new and do reveal an 
interplay with the various versions of duality.

Fluxes of the electromagnetic field can only occur if $\M_4$ possesses 
nontrivial two-cycles. Roughly these are closed surfaces through which
the flux may flow. Topologically they may be spheres, two-tori, surfaces
of any genus or even integer linear combinations of these.

Because $\M_4$ is oriented, such two-cycles $\Sigma$ can be assigned
an integer $I(\Sigma,\Sigma)$ counting the number of points of
self-intersection taking into account orientation. To understand 
this idea it is necessary to imagine a small distortion of $\Sigma$
away from itself to form $\Sigma'$. Then, if this is done appropriately,
the two surfaces $\Sigma$ and $\Sigma'$ intersect in a finite number
of points and the resultant self-intersection number is independent
of how this is done.

Flux quantisation is a quantum phenomenon, of course. Accordingly one 
considers a complex scalar wave function for a bosonic particle carrying 
electric charge $q_B$. Given its existence on $\M_4$, Dirac's
condition [1931] reads
$$
{q_B\over2\pi\hbar}\int_{\Sigma}F\in\Zs.
\eqno(1.1)
$$
Now consider a complex spinor wave function for a fermionic particle 
carrying electric charge $q_F$. Our result is that, instead,
$$
{q_F\over2\pi\hbar}\int_{\Sigma}F+{1\over 2} I(\Sigma,\Sigma)\in\Zs.
\eqno(1.2)
$$
Notice that this condition, unlike (1.1) is special to four-manifolds
as the intersection number of two-cycles only makes sense there.
Notice also that if the four-manifold $\M_4$ possesses any two-cycle 
with odd self-intersection number the flux through it must be fractional.
In particular the flux cannot vanish. Putting $q_F$ to zero in (1.2)
then yields a contradiction. This means that such manifolds cannot
support neutral  spinors. The simplest example of this, already
known for some time, is when $\M_4$ is the complex projective
space $CP(2)$ [Hawking and Pope 1978].

Likewise, if $\M_4$ does possess real (uncharged) spinor wavefunctions 
all two-cycles must have even self-intersection number. This was first
concluded on the basis of totally different arguments by Geroch [1968 and
1970]. But the converse  is not true. There do exist four-manifolds, all of
whose two-cycles   possess even self-intersection number, that do not support
a real spinor field. All these four four-manifolds have the feature that   
they cannot be simply connected. The simplest example is the product of
two two-spheres divided by a parity $Z_2$ [Habegger 1982, Kirby 1989].

Our treatment explains these phenomena in a unified way. It uses heavily the
feature that the free two-cycles on $\M_4$ form a unimodular lattice and
that the permitted fluxes are related to a similar lattice (the reciprocal).
Such unimodular lattices possess special elements, known as characteristic
vectors. Their components are related to the self-intersection numbers
occurring in the condition (1.2) in a way that is explained, based an
application of the Atiyah-Singer index theorem for the electromagnetic Dirac
operator on $\M_4$. The unimodularity feature is an essential ingredient
in verifying electromagnetic duality as the quantum Maxwell partition
function turns out to be a generalised theta function based upon a sum
over this lattice [E Verlinde 1995, E Witten 1995].

An important role in our arguments is played by various versions of
what we call a ``quantum Stokes' theorem''. Ordinary Stokes' theorem
does not apply to gauge potentials on topologically complicated
manifolds because the potential is only defined in contractible
regions. To deal with this the manifold must be covered with
contractible neighbourhoods. Then the gauge potentials in overlapping
neighbourhoods must be patched together with gauge
transformations. This procedure introduces ambiguities in the
candidate Stokes' theorem which can be eliminated by a suitable
exponentiation, yielding the exponential version of Stokes' theorem
that we refer to as quantum Stokes' theorem.

The success of this procedure presupposes that the underlying complex
wave function is scalar. But when the complex wave function
is spinor rather than scalar, the exponentiated line integral acquires an extra
overall sign ambiguity stemming from the double-valued nature of the spinor
representation. This sign can be eliminated only by squaring it. This
phenomenon means that, in the presence of complex spinor wave functions, the
quantum Stokes' theorem can be formulated only for fluxes through surfaces
whose boundary either vanishes or is even (that is twice a closed curve). Such
surfaces are respectively orientable and non-orientable. There remains an
unexpected sign but this is now well-defined and unambiguous, taking the
values $\pm1$ according to whether what is called the Stiefel-Whitney class of
the surface is even or odd (see equation (6.4)). 

This result and its implications match precisely with a construction
known to mathematicians from a different point of view, and referred
to as $Z_2$ cohomology. Thus we shed a more physical light on this
mathematical nicety together with the Stiefel-Whitney class which is
an element of the second $Z_2$ cohomology group. Following from this
is  the distinction between the quantisation conditions (1.1) and
(1.2) and other more subtle physical effects that we discuss. The
Stiefel-Whitney class is shown to be a concept extending that of the
self-intersection number in (1.2), and equivalently, the
characteristic vector of the flux lattice. This is encapsulated in a
formula known as Wu's formula.

The reader will notice that we have never used the words ``magnetic
monopole''. This is because there are none, as far as the space-time
$\M_4$ is concerned. One can imagine that the fluxes originate with
magnetic monopoles but these have to be situated somewhere outside
space-time. This means there is never any question of singularities in the fields at points of $\M_4$ corresponding to their positions.
\bigskip
\ni {\bf 2. Four-manifolds and their homology}
\medskip
We are going to be interested in space-times consisting of compact,
connected and oriented four-manifolds $\M_4$. This is because such
manifolds  obey Poincar\'e duality, a symmetry of the topological
properties that seems to be related to the electromagnetic duality
that is our main interest. We shall not require $\M_4$ to be simply
connected as this would exclude one of the most interesting
phenomena. The sort of topological structure of $\M_4$ that is
relevant to the study of electromagnetic fluxes is known as homology
(and cohomology) and is described in many textbooks to which we refer,
\eg [A Schwarz 1994].

Thus $H_p(\M,\Zs)$ denotes the abelian  group formed by adding 
(over integers $\Zs$) classes of closed $p-$cycles mod $p-$boundaries.
Dual to the boundary operator, $\partial$, is the coboundary operator,
$\delta$, which likewise squares to zero. It defines the cohomology
groups for coclosed $p-$cocycles mod $p-$coboundaries, $H^p(\M,\Zs)$.
We shall be supposing $\M$ to be sufficiently well behaved that
these groups be finitely generated. Each may contain elements of
finite order. These form well-defined finite subgroups, called
the torsion subgroups, denoted $T_p(\M,\Zs)$ and $T^p(\M,\Zs)$,
respectively. Thus a $p-$cycle $\alpha$ has finite order $N$ 
if $N$ is the smallest integer such that $N\alpha$ is a boundary:
$$
N\alpha=\partial\beta.
\eqno(2.1)
$$
As $\partial^2=0$, $\partial\alpha=0$, that is, $\alpha$ is closed.

The torsion subgroups are automatically selfconjugate subgroups 
and the quotients
$$
F_p(\M,\Zs)=H_p(\M,\Zs)/T_p(\M,\Zs),\qquad
F^p(\M,\Zs)=H^p(\M,\Zs)/T^p(\M,\Zs)
\eqno (2.2)
$$
are infinite free abelian groups taking the form of a finite 
number of copies of the integers $\Zs$. Thus
$$
F_p(\M,\Zs)=\Zs\oplus\Zs\oplus\dots \Zs=\oplus_{b_p}\Zs.
\eqno(2.3)
$$
The integers $b_p$ and $b^p$ counting these numbers of copies are known 
as the $p-$th Betti numbers of $\M$. If one works over real numbers 
(instead of integers) the torsion becomes invisible, leaving just the
free parts. It will also be relevant to work over the integers 
mod $2$, \ie $Z_2$, but the corresponding groups $H_p(\M,Z_2)$ 
and $H^p(\M,Z_2)$ contain no new information since they can be 
constructed explicitly from the $H_p(\M,\Zs)$ and $H^p(\M,\Zs)$ 
by what is known as the universal coefficient theorem. This theorem 
also implies the following isomorphisms between cohomology and homology:
$$
F^p(\M,\Zs)\equiv F_p(\M,\Zs),\qquad T^p(\M,\Zs)\equiv
T_{p-1}(\M,\Zs).
\eqno(2.4)
$$
In particular, the Betti numbers $b_p$ and $b^p$ are equal. 
So far everything has been rather general but now the special
fact that $\M$ is assumed to possess the features guaranteeing 
Poincar\'e duality means that:
$$
H^p(\M,\Zs)\equiv H_{m-p}(\M,\Zs),
\eqno(2.5)
$$
where $m$ is the dimension of $\M$, (understood to be $4$ when $\M_4$
is indicated). In particular, yet more Betti numbers become equal:
$$
b_p=b^p=b_{m-p}=b^{m-p}.
\eqno(2.6)
$$
Cycles $\Sigma$ and $\Sigma'$ of complementary dimensions (\ie adding 
to $m$) usually intersect in a finite number of discrete points. Owing
to the fact that $\M$ is oriented these points can be assigned values
$\pm1$ and the sum of these values over the points of intersection
yields an integer known as the intersection number of $\Sigma$ and $\Sigma'$,
denoted $I(\Sigma,\Sigma')$. This integer only depends on the homology
classes to which the cycles belong, and more than that, only upon the
cosets $F_p(\M,\Zs)$ to which they belong. This is because all torsion cycles have zero intersection number with any other cycle. For
$$
I(\alpha,\gamma)=
I(N\alpha,\gamma)/N=I(\partial\beta,\gamma)/N=I(0,\gamma)/N=0,
$$
using linearity and equation (2.1).

On four-manifolds, $2-$cycles will have such discrete points of
intersection with each other. Let us consider a basis
$\Sigma_1,\Sigma_2,\dots \Sigma_{b_2}$ of the integer lattice
$F_2(\M_4,\Zs)$ and denote
$$
I(\Sigma_i,\Sigma_j)=(Q^{-1})_{ij}.
\eqno(2.7)
$$
$Q$ is a matrix with $b_2$ rows and columns and integer entries. It is
unimodular \ie has determinant $\pm1$, another consequence of
Poincar\'e duality. Finally it is symmetric (as it is for any such
intersection matrix for mid-cycles when $m$ is a multiple of four):
$$
Q=Q^T\in\Zs,\qquad \det \,Q=\pm1.
\eqno(2.8)
$$
This matrix and these properties will play an important role in what
follows. We see that the free $2-$cycles form a unimodular lattice,
namely $F_2(\M_4,\Zs)$. Notice that the scalar product of this
lattice, as defined by the matrix $Q^{-1}$, is, in general,
indefinite. If it is of type $(b^+,b^-)$ where $b_2=b^++b^-$ the signature
$$
\sigma(\M_4)=b^+-b^-
\eqno(2.9)
$$ 
is known as the Hirzebruch signature of $\M_4$.
\bigskip
\ni{\bf 3. Fluxes and the Dirac quantisation condition}
\medskip
Let $F$ be a two-form field strength defined on $\M$. It is closed:
$$
dF=0,
\eqno(3.1a)
$$ and as a consequence there exists a one-form $U(1)$ connection $A$ such that
$$
F=dA
\eqno(3.1b)
$$
at least in each contractible neighbourhood used in a covering of
$\M$. Associated with any $2-$cycle $\Sigma$ is a flux
$$
\int_{\Sigma}F.
\eqno(3.2)
$$
Stokes theorem guarantees that this flux is unaltered if we replace
$\Sigma\rightarrow\Sigma'=\Sigma+\partial R$ and $F\rightarrow
F'=F+dB$. Thus the flux (3.2) depends only on the homology class of
$\Sigma$ (or rather the coset $F_2(\M_4,\Zs)$) and the cohomology
class of $F$. However there is a subtlety as the cohomology used here
is known as de Rham cohomology and works with real
coefficients. Integer coefficients were restored by Dirac [1931] by
showing that if the gauge potential $A$ coupled minimally to the
scalar wave function defined on space-time, then the principles of
quantum mechanics requires a quantisation of the flux
$$
q_B\int_{\Sigma}{F\over2\pi\hbar}\in \Zs
\eqno(3.3)
$$
where $q_B$ is the electric charge carried by the bosonic particle 
created by the scalar field when quantised. That this equation (3.3) 
is independent of any assumed equations of motion and of any choice
of metric on $\M$ was made particularly clear in a version of this
argument due to Wu and Yang [1975]. However this worked only when
the $2-$cycle $\Sigma$ was topologically a sphere. The extension 
to any $2-$cycle was explained by O Alvarez [1985] and we shall
extend his version further in what follows. 

Before doing so, let us consider the integral $\int_{\M_4}F\wedge F$ 
which occurs both as part of the Maxwell action and the Atiyah-Singer 
index for the electromagnetic Dirac equation on $\M_4$. Again both $\M_4$ 
and $F\wedge F$ are closed so a definite value can be anticipated. 
In fact the following identity holds
$$
\int_{\M_4}F\wedge F=\sum_{i,j=1}^{b_2}\int_{\Sigma_i}
F\,\,\,Q_{ij}\,\,\int_{\Sigma_j} F.
\eqno(3.4)
$$
This is just a special case of a much more general identity which,
when applied to Riemann surfaces yields Riemann's bilinear identity. 
$Q$ is defined in (2.7) and it is now clear from (3.4) why it should
be symmetric. Since fluxes through the torsion cycles always vanish 
(by a similar argument to that above for intersection numbers),
it is entirely equivalent to rewrite the Dirac quantisation 
condition (3.3) in the form
$$
q_B\int_{\Sigma_i}{F\over2\pi\hbar}=m_i\in \Zs.
\eqno(3.5)
$$
Inserting this into (3.4) yields
$$
q_B^2\int_{\M}{F\over2\pi\hbar}\wedge{F\over2\pi\hbar}=\sum_{i,j=1}^{b_2}m_iQ_{ij}m_j.
\eqno(3.6)
$$ 
Thus this integral is also quantised, its integral value being
quadratic in the quantised fluxes and depending on the topology 
of $\M$ through the intersection matrix $Q$. Physicists are used
to this integral for non-abelian gauge theories when it is the 
instanton number and exists even if $\M$ is the sphere $S_4$. 
But (3.6) vanishes whenever $b_2$ vanishes, for example for
$S_4$ or $R_4$. On the other hand, the second Betti number 
is non-zero for many interesting four-manifolds \eg $b_2(T^4)=6$,
$b_2(CP(2))=1$ and $b_2(K_3)=22$.

Note that (3.6) means that the Dirac quantised fluxes on $\M$ form a unimodular
lattice (because the scalar product $Q$ has unit determinant) with
signature (2.9). Such lattices have interesting properties [see
Goddard and Olive 1984] and can be used in the construction of functions
transforming nicely under an action of the modular group.

Now recall the argument of O Alvarez [1985]. $\M_4$ is covered with
a finite number of neighbourhoods $\ua$. Since $\ua$ is contractible, 
there exists in it a $U(1)$ gauge potential/connection $\aa$ and a scalar
wave function $\pa$. In the overlap region $\uab$ of two contiguous 
neighbourhoods $\ua$ and $\ub$, also assumed to be contractible, these
are related by a $U(1)$ gauge transformation:
$$
\aa-\ab=d\cab, \qquad \pa=e^{\im{q_B\over\hbar}\cab}\pb.
\eqno(3.7)
$$
Without loss of generality we can assume $\cab+\cba=0$ in $\uab$. In
the triple overlap region of three contiguous neighbourhoods, $\uabc$,
all three of $\pa$, $\pb$ and $\pc$ are all defined and related by the
three gauge transformations $\cab$, $\cbc$ and $\cca$. The self
consistency of these three actions on the scalar field requires that
$$
c^{\alpha\beta\gamma}\equiv\cab(P)+\cbc(P)+\cca(P)\in {2\pi\hbar\over q_B}\Zs,
\eqno(3.8)
$$
for any point $P$ in $\uabc$. So $c^{\alpha\beta\gamma}$ is
independent of the choice of this point $P$.

Now apply this to a $2-$cycle $\Sigma$ and suppose that it is covered
by a finite number of neighborhoods $\ua\cap\Sigma$ with no more than triple
overlaps. Further imagine that these neighbourhoods $\ua\cap\Sigma$ 
be reduced to non overlapping neighbourhoods $\va$ sharing a common 
boundary when $\ua$ and $\ub$ are contiguous. Then the flux of $F$ 
through $\Sigma$ is a sum of the fluxes through the $\va$, since $F$ 
is gauge invariant and so $F^{\alpha}=F^{\beta}$ in $\uab$. 
Ordinary Stokes' theorem applies to each $\va$ so
$$
\int_{\Sigma}F=
\sum_{\alpha}\int_{\va}F=\sum_{\alpha}\int_{\partial\va}A^{\alpha}.
$$
Each common boundary $\lab$ of $\va$ and $\vb$ in $\uab$ contributes 
a line integral $\int_{\lab}(\aa-\ab)=\cab|_{\partial\lab}$. Each 
triple overlap $\uabc\cap\Sigma$ has three such boundary lines incident
on it and, as a result, contributes $c^{\alpha\beta\gamma}$ to
$\int_{\Sigma}F$. Hence
$$
\int_{\Sigma}F=\sum_{\uabc\cap\Sigma}\,\,\,c^{\alpha\beta\gamma}\in
{2\pi\hbar\over q_B}\Zs
$$ 
by (3.8). This is the Dirac quantisation condition (3.3).

More can be extracted from this line of argument by considering 
first a closed $1-$cycle $\gamma$ and seeking to define the line 
integral of $A$ around it. Suppose $\gamma$ is contained in a sequence
of consecutive neighbourhoods $\U_1,\U_2,\dots \U_N=\U_0, \U_{N+1}=\U_1$. 
Let $P_{k,k+1}$ be a point on $\gamma\cap \U_k\cap \U_{k+1}$ and
break $\gamma$ into consecutive segments
$P_{1,2}P_{2,3},P_{2,3}P_{3,4},\dots ,P_{k,k+1}P_{k+1,k+2},\dots$. Then a
sensible definition might be
$$
{\hbox{``}}\oint_{\gamma}A{\hbox{''}} =\sum_k\Big(\int_{P_{k-1,k}}^{P_{k,k+1}}
A^k-\chi^{k,k+1}(P_{k,k+1})\Big).
\eqno(3.9)
$$ 
This has the virtue of being independent of the choices of  the points
$P_{k,k+1}$, as can be seen by differentiation with respect to
them. But it is not independent of the choice of neighbourhoods on
$\gamma$. Indeed adding a neighbourhood modifies the above sum to
include an extra term but also adds a term of the form (3.8). Thus the
line integral has an quantised ambiguity of the form of integral multiples of
$2\pi\hbar/q_B$. This ambiguity can then be eliminated by considering
only the exponential $\exp(\im q_B\oint_{\gamma}A/\hbar)$, with the
exponent defined by (3.9). Thus it is only possible to define a line
integral of a $U(1)$ gauge potential on a non-trivial manifold in this way.

If now $\gamma$ is a boundary $\partial \Sigma$, the argument can be
repeated to give 
$$
\int_{\Sigma}F={\hbox{``}}\oint_{\gamma}A{\hbox{''}} 
+\sum_{\uabc\cap\Sigma}\,\,\,c^{\alpha\beta\gamma}.
\eqno(3.10)
$$
Again because the ambiguities are quantised they can be eliminated by
exponentiation: 
$$ \e^{\im{q_B\over\hbar}\int_{\Sigma}F}=
\e^{\im{q_B\over\hbar}\oint_{\gamma}A}.
\eqno(3.11)
$$ 
If $\gamma$ contracts to a point (3.11) reduces to the Dirac
quantisation condition (3.3) or (1.1).

Because of the essential way that Planck's constant occurs in (3.10),
we can think of (3.11) as a ``quantum version'' of Stokes'
theorem. Notice that without the notion of quantum wave function 
it would be difficult to formulate any version of Stokes' theorem 
for $U(1)$ gauge potentials on non-trivial manifolds. It is entirely natural
that the right hand side of (3.11) should be Dirac's path dependent
phase factor, nowadays known as a ``Wilson loop''.

\bigskip
\ni{\bf 4. Spinor fields and the Dirac quantisation condition}
\medskip
The preceding argument shows how dependent the behaviour of the fluxes
on $\M$ is upon the assumed existence of the complex scalar field
$\phi$ on $\M$. It is natural therefore to examine what happens in the
presence of a complex spinor field $\psi$. It is known that a real
(neutral) spinor field need not necessarily exist on $\M$. There
exists a possible obstruction, known as the Stiefel-Whitney class
$w^{(2)}$ which is an element of $H^2(\M,Z_2)$. Its vanishing, mod 2,
is the necessary and sufficient condition for neutral spinors to exist
on $\M$.In this case $\M$ is said to possess a ``spin
structure''. If it does not vanish it may be possible to define a
complex rather than a real spinor field on $\M$. Such a field, $\psi$,
would carry an electric charge $q_F$, thereby coupling to $A$, rather 
as the complex scalar field $\phi$ did (but with charge $q_B$).

In fact it is known that this is possible whenever $w^{(2)}$ ``lifts''
from $H^2(\M,Z_2)$ to an element of $H^2(\M,\Zs)$ when integer
coefficients are used. According to the universal coefficient theorem,
the $w^{(2)}$ can lift either to $H^2(\M,\Zs)$ or to
$T^3(\M,\Zs)$. These  possibilities are easy to understand. If $\delta
w^{(2)}=0$, mod two, then, when integer coefficients are considered,
either $\delta w^{(2)}=0$ (and we have an element of $H^2(\M,\Zs)$) or
$\delta w^{(2)}=2\lambda$ where $\lambda$ is a $3-$co-chain. As
$\delta$ squares to zero it follows that $\lambda$ is a
$3-$torsion-cocycle, thereby furnishing an element of $T^3(\M,\Zs)$ of
order two.

When $w^{(2)}$ lifts to $H^2(\M,\Zs)$ a complex spinor can exist on
$\M$ and $\M$ is said to possess a ``spin$_C$'' structure. But
reasonable counterexamples exist whenever the dimension of $\M$, $m$,
is five or more. For example, the five-dimensional symmetric space
$SU(3)/SO(3)$ has no spin$_C$ structure [Lawson and Michelsohn 1989].

However for four-manifolds with Poincar\'e duality something special happens.
The conditions (2.4) and (2.5) together with the statement that $\M_4$
is connected imply that there are just two independent Betti numbers,
$b_1=b_3$and $b_2$ as $b_0=b_4=1$, and one independent torsion group
because of the isomorphisms 
$$
T^3(\M_4,\Zs)=T^2(\M_4,\Zs)=T_2(\M_4,\Zs)=T_1(\M_4,\Zs).
\eqno(4.1)
$$ 
The other torsion groups vanish.

We can now see what an undesirable degree of simplification would 
occur were it assumed, in addition, that $\M_4$ was simply connected.
Then $H_1(\M_4,\Zs)$ would vanish, and hence, by (4.1), all the
torsion groups.

On the four-manifolds considered it can always be said that $w^{(2)}$ 
lifts to $H^2(\M_4,\Zs)$ as $T^3$ and $T^2$ are the same thing, according
to the isomorphisms (4.1). So such four-manifolds always possess
spin$_C$ structures and hence complex spinor fields.

Accordingly, we shall henceforth assume that complex spinor fields
exist on $\M_4$ and consider the effect on Dirac's quantisation 
condition. So we reconsider the argument of the previous section 
and take the same finite cover of $\M_4$ by neighbourhoods $\ua$. 

Because the wave functions now transform as $so(4)$ spinors we have
to introduce more structure in each neighbourhood. As well as the
previous $U(1)$ connection or gauge potential $\aa$ we must introduce
an oriented frame (vierbein) $\fa$ and a complex spinor wave-function
$\psi^{\alpha}$. These choices, made in the neighbourhood $\ua$, are
not independent. There is a freedom of choice related to local $U(1)$
gauge transformations 
$$
\eqalign{
\psi^{\alpha}&\rightarrow\e^{\im{q_F\over\hbar}\chi}\psi^{\alpha},\cr
\aa&\rightarrow\aa+d\chi,\cr}
$$
and to local $so(4)$ transformations
$$
\eqalign{
\fa&\rightarrow L\,\fa,\cr
\psi^{\alpha}&\rightarrow S(L)\psi^{\alpha},\cr}
$$
where $L\in so(4)$. Note that the lift $L\rightarrow\pm S(L)$ has a
sign ambiguity because the quotient of the spin group $spin(4)$ by $Z_2$ is 
isomorphic to $so(4)$.

In a double overlap of two contiguous neighbourhoods, $\uab\neq0$, 
we must define transition functions involving both transformation groups
$$
\eqalignno{
\aa&=\ab+d\cab,&(4.2a)\cr
F^{\alpha}&=\lab F^{\beta},&(4.2b)\cr
\psi^{\alpha}&=S(\lab)\,\e^{\im{q_F\over\hbar}\cab}\psi^{\beta}.&(4.2c)\cr}
$$
Without loss of generality, we can suppose 
$$
\lab=(L^{\beta\alpha})^{-1},\qquad
\cab=-\chi^{\beta\alpha}\quad\hbox{and}
\quad S(\lab)=(S(L^{\beta\alpha}))^{-1}.
\eqno(4.3)
$$
In a triple overlap region, $\uabc\neq0$, again assumed to be contractible,
we find consistency conditions as before;
$$
\lab\lbc\lca=I,
\eqno(4.4)
$$
where the right hand side is the unit element of $so(4)$. This follows
by relating the frame $\fa$ back to itself. Likewise for the spinor
wave-function 
$$
\psi^{\alpha}=\e^{\im{q_F\over\hbar}(\cab+\cbc+\cca)}S(\lab)S(\lbc)S(\lca)
\psi^{\alpha}.
$$
The product of the three matrices $S$ here will play an important role
in what follows and we shall introduce a special notation for that product
$$
\eabc\equiv S(\lab)S(\lbc)S(\lca)=\pm I.
\eqno(4.5)
$$
That the right hand side is the unit matrix, up to a sign, follows by 
lifting (4.4) from $so(4)$ to $spin(4)$. The sign cannot be determined
as it depends on the choices made in (4.2). Nevertheless the signs in
(4.5) for different triple overlaps are not totally independent as we
shall see. So the upshot of the spinor consistency condition is that,
in terms of $c^{\alpha\beta\gamma}$ defined in (3.8),
$$
\e^{\im{q_F\over\hbar}c^{\alpha\beta\gamma}}=\eabc,
\eqno(4.6)
$$
rather than unity, as was the case for the scalar wave function 
(with $q_B$ replacing $q_F$). So, repeating the argument for 
evaluating the flux through a two-cycle $\Sigma$, we find
$$
\e^{\im{q_F\over\hbar}\int_{\Sigma}F}=(-1)^{w(\Sigma)},
\eqno(4.7)
$$
where the sign is given by the finite product over triple overlaps
$$
(-1)^{w(\Sigma)}=\prod_{\uabc\cap\Sigma\neq0}\eabc.
\eqno(4.8)
$$
Equation (4.7) is the the preliminary version of the Dirac 
quantisation condition in the presence of a complex spinor wave-function.
Notice how $w(\Sigma)$ (which is defined mod $2$) must be independent 
of the choices made in covering the two-cycle $\Sigma$ with neighbourhoods,
even though its definition (4.8) involved those choices. 
This is because of its relation (4.7) to the flux which is certainly
independent of those choices. Likewise $w(\Sigma)$ is unchanged 
if the two-cycle is replaced by a homologous one 
$\Sigma\rightarrow\Sigma'=\Sigma+\partial R$. Furthermore $w(\Sigma)$
vanishes if $\Sigma$ is a torsion element. All this suggests that $w$ is 
closely related to the Stiefel-Whitney two-cocycle $w^{(2)}$ over
$Z_2$ mentioned earlier, and this is indeed true as we shall see. 

Denoting $w_i=w(\Sigma_i)$, where ${\Sigma_i}$ is the basis of
$F_2(\M_4,\Zs)$ introduced in section 2, the quantisation condition
can be rewritten as
$$
{q_F\over2\pi\hbar}\int_{\Sigma_i}F-{1\over 2} w_i\in \Zs,\qquad
i=1,2,\dots,b_2.
\eqno(4.9)
$$
So we see the possibility of fractional units of flux. We still have
to identify the number $w_i$ with the self-intersection number of
$\Sigma_i$ mod $2$ as described in the introduction.

In this context the possibility of fractional fluxes was first pointed
out by Hawking and Pope [1978] who considered just the example of
$\M_4$ being $CP(2)$. However a similar effect had already been
established in the context of gauge theories in $R^{3,1}$ describing
both colour and a Maxwell $U(1)$ [Corrigan and Olive 1976]. This
modified Dirac quantisation condition could explain how the fractional
electric charge of quarks was related to their colour transformation
properties.

\bigskip
\ni {\bf 5. Atiyah-Singer index theorem and identification 
of $w(\Sigma)$}
\medskip
Because  complex spinors always exist on any $\M_4$ under
consideration, we can evaluate the index of the Dirac operator $D_A$
acting on them. According to the index theorem the result is
$$
\Index(\dir)=-{1\over 8}\,\sigma(\M_4)+{q_F^2\over8\pi^2\hbar^2}
\int_{\M_4}F\wedge F
\eqno(5.1)
$$
Such results are familiar to physicists because of their relation to
the chiral anomaly.

We have already met the expressions on the right hand side. The
signature $\sigma$ is that given by (2.9) while the integral is
evaluated by (3.4) making use of the Dirac quantisation conditions in
the version (4.9). The result is 
$$
\eqalign{
\Index(\dir)&=
-{1\over 8}\,\sigma(\M_4)+{1\over 2}(m+{w\over2})\,Q\,(m+{w\over2})\cr
&={1\over 8}(w\,Q\,w-\sigma(\M_4))+{1\over2}(m\,Q\,m+m\,Q\,w),\cr}
\eqno(5.2)
$$
on rearrangement. The quantities $m_1,m_2,\dots ,m_{b_2}$ are
integers. Because the spinors exist on $\M_4$, this has to be an
integer for all integers $m_i$. As a result we deduce the following
two conditions 
$$
w\,Q\,w=\sigma(\M_4)+8\Zs
\eqno(5.3a)
$$
$$
m\,Q\,m+m\,Q\,w=2\Zs.
\eqno(5.3b)
$$
Recall that the matrix $Q$ is integer valued and unimodular with
signature $\sigma(\M_4)$. There is a general theorem that states that
such a matrix possesses a characteristic vector $c$, say, satisfying
(5.3b) [Milnor and Husemoller 1973]. This vector is unique up to an
ambiguity in its components of precisely $2\Zs$. Furthermore it
automatically satisfies (5.3a). In other words, equation (5.3b)
completely determines $w_i$ mod $2$ in terms of the intersection
matrix $Q^{-1}$. In fact, inserting into (5.3b) the integral choices  
$m_i=(Q^{-1})_{ik}$, ($k=1,2,\dots$ or $b_2$) we find the solution
$$
w_k=-Q^{-1}_{kk}+2\Zs=-I(\Sigma_k,\Sigma_k)+2\Zs,
\eqno(5.4)
$$
using (2.7) for the self-intersection number explained in the introduction.

Unimodular (integral) matrices are said to  be  even if all diagonal 
elements $Q_{ii}$ (or, equivalently all $Q_{ii}^{-1}$) are even. 
Otherwise they are  odd. In the former case $w_i$ is even by (5.4), 
and so can be taken to vanish. In that case (5.3a) states that the 
Hirzebruch signature is a multiple of eight. On the other hand, 
if $Q$ is odd, $w_i$ is non vanishing for at least one value of
the suffix $i$.

Thus the quantisation condition (1.2) has been established for all the
$\Sigma_i$. As these form a basis for the free $2-$cycles  
$F_2(\M_4,\Zs)$, the validity of the quantisation condition (1.2)
extends to all of these. We have already seen that both the 
flux and the self-intersection number vanish for torsion
$2-$cycles. Hence the result (1.2) is established for all
$2-$cycles $\Sigma$.

It is now easy to understand the example of $\M_4$ being complex
projective space $CP(2)$ [Hawking and Pope 1978]. As the Betti number
$b_2$ equals one, $Q=\pm1$, by unimodularity and hence is odd. So
$CP(2)$ possesses only one $2-$cycle (actually a sphere) and that has
odd self-intersection. As a result there is no spin structure but
complex spinors are allowed. As the Hirzebruch signature cannot vanish
(mod $8$) when the second Betti number is odd, the concluded absence
of real spinors extends to any four-manifold with odd second Betti number.
\bigskip
\ni{\bf 6. New aspects of the Dirac quantisation condition}
\medskip
In section 3, we saw how a `` quantum Stokes' theorem'' could be
established, (3.11), in connection with a complex scalar wave
function. It is natural to enquire about an analogous result in
connection with a complex spinor wave function. In general, it is
bound to differ as we have already seen that it does when a two-cycle
with odd self-intersection number is considered.

In order to continue the argument of section 4 to the case that
$\Sigma$ has a boundary we need to understand better how it is that
the sign factor $(-1)^{w(\Sigma)}$ defined by (4.8) is independent of
the choice of covering used in this definition. To this end we must
investigate the properties of the sign factors $\eabc$ defined by
(4.5) and associated with triple overlaps. 

The first point to note is that $\eabc$ is totally symmetric in its
three indices. For $\eabc$ is conjugate to $\epsilon(\beta\gamma\alpha)$. 
But since it is proportional to the unit matrix it commutes with all 
the matrices $S(\lab)$. So $\eabc$ is invariant with respect to cyclic
permutations. It remains to check invariance under anticyclic
permutations and this follows from the identity 
$\eabc\epsilon(\gamma\beta\alpha)=1$, proven only using (4.3).

Now let us suppose that there is a quadruple overlap of four
neighbourhoods $\ua$, $\ub$, $\uc$ and $\U_{\delta}$. In it, four
quantities (4.5) can be defined, but not independently, as their
product equals the unit matrix. 
$$
\eabc\epsilon(\beta\gamma\delta)\epsilon(\gamma\delta\alpha)
\epsilon(\delta\alpha\beta)=I.
\eqno(6.1)
$$
In the quadruple overlap we can also define new quantities with four
indices, for example: 
$$
\epsilon(\alpha\beta\gamma\delta)
\equiv S(\lab)S(\lbc)S(L^{\gamma\delta})S(L^{\delta\alpha})
$$
which also equals the unit matrix up to a sign as it is the lift of
$$
\lab\lbc L^{\gamma\delta}L^{\delta\alpha}=I
$$
These quantities are likewise invariant under cyclic and anticyclic
permutations of their four indices. That is no longer sufficient to
guarantee  complete symmetry and it leaves three distinct quantities
specified by signs when the indices are distinct. Again using only (4.3)
$$
\epsilon(\beta\gamma\alpha)\epsilon(\beta\alpha\delta)=
\epsilon(\beta\gamma\alpha\delta).
\eqno(6.2)
$$
There are six identities of this type and they imply (6.1).

We can now use (6.1) to show how the addition of a neighbourhood 
$\U_{\delta}$, say to a cover of $\Sigma$ does not affect
$(-1)^{w(\Sigma)}$ given by (4.8) provided it affects only the
interior. We shall compare the original covering without $\U_{\delta}$
to the one in which $\U_{\delta}$ is added so that it includes the
triple overlap $\uabc$. We can then imagine that $\ua$, $\ub$ and $\uc$ are 
contracted in such a way that $\uabc$ disappears. The effect is that
the triple overlap $\uabc$ is replaced by three triple overlaps 
$\ua\cap\ub\cap\U_{\delta}$, $\ub\cap\uc\cap\U_{\delta}$ and 
$\uc\cap\ua\cap\U_{\delta}$. So, in $(-1)^{w(\Sigma)}$, $\eabc$ is
replaced by $\epsilon(\beta\gamma\delta)\epsilon(\delta\gamma\alpha)
\epsilon(\alpha\gamma\delta)$. But these are the same by (6.1), as required.

Now consider the flux through $\Sigma$ with non-trivial boundary 
$\partial\Sigma$. The argument leading to (3.10) still holds good and
so the equation remains valid. The difference is that now 
$c^{\alpha\beta\gamma}$ satisfies (4.6) and so the exponential could be minus 
one instead of just plus one. That means that exponentiation yields
$$ 
\e^{\im{q_F\over\hbar}\int_{\Sigma}F}= \e^{\im{q_F\over\hbar}
\oint_{\partial\Sigma}A}\prod_{\uabc\cap\Sigma\neq0}\eabc.
\eqno(6.3)
$$
The problem with this result is that although the right hand side is
independent of the choice of neighbourhoods the two individual factors are not.
We have just argued that adding a neighbourhood to the interior of
$\Sigma$ does not affect the second factor and obviously it does not
affect the first factor. The problem comes when we add a neighbourhood to the 
cover of the boundary $\partial\Sigma$. That can change the sign of
both factors. This means that the individual factors are not
intrinsically defined and we do not know of any better way of
formulating Stokes' theorem in this context. Of course we could
eliminate the problem by squaring (6.3). This would yield the same
result as (3.11) with $q_B$ replaced by $2q_F$ but this discards information.

However there is a situation where (6.3) is useful. Suppose that the
boundary of $\Sigma$ is even, that is that
$\partial\Sigma=2\alpha$. Then we do have 
$$
\e^{\im{q_F\over\hbar}\int_{\Sigma}F}
=(-1)^{w(\Sigma)}e^{2\im{q_F\over\hbar}\oint_{\alpha}A}
\eqno(6.4)
$$
and the two factors on the right hand side are well defined in the
sense that they are independent of choices of neighbourhoods. This is
because the second factor is, in view of the factor two in the
exponent. It follows that $w(\Sigma)$ is well defined (mod $2$) on
such $\Sigma$ with even boundary.

We shall argue that there is good reason to regard (6.4) as a new
aspect of the Dirac quantisation condition. 

Suppose the four-manifold $\M_4$ is equipped with a spinor and that
all its two-cycles possess even self-intersection number. That means,
as we have shown, that all fluxes are integral and in particular may
all vanish. Then it is possible and reasonable to take vanishing
field strength on $\M_4$. Then, one might think that the spinor experiences no
electromagnetic coupling so that it could perfectly well be taken to
be real. But this conclusion can be wrong, as (6.4) shows. For,
suppose $\M_4$ contains a two-chain $\Sigma$ whose boundary is even,
that is twice a one-cycle $\alpha$, as in (6.4). Suppose further that
$w(\Sigma)$ is non-vanishing (mod $2$). Then (6.4) reads 
$$
e^{2\im{q_F\over\hbar}\oint_{\alpha}A}=-1.
$$
This means that it is impossible for $q_F$ (or $A$) to vanish. In
other words we have a new sort of obstruction to the existence of real
spinors. Crudely speaking such four-manifolds possess fluxes leaking
through holes enclosed by torsion cycles such as $\alpha$, even when
the field strength vanishes on the manifold itself. The situation
$\M_4=S^2\times S^2/Z_2$ mentioned in the introduction furnishes an
example of precisely this. In this case $\Sigma$ is topologically the real projective plane and so not orientable, and this is  a general feature.

There is a construction, known to mathematicians as Wu's formula, for
the action of the Stiefel-Whitney cocycle acting on two-cycles when it
is understood that in all calculations integers are identified mod $2$. It is
$$
w^{(2)}(\Sigma)=J(\Sigma,\Sigma)\quad \hbox{mod}\,\, 2,\quad  
\Sigma\in H_2(\M_4,Z_2).
\eqno(6.5)
$$
Here $J(\Sigma,\Sigma')$ is the intersection number of $\Sigma$ and
$\Sigma'$, defined now as simply the number of points of intersection
mod $2$. This definition, unlike the one for $I(\Sigma,\Sigma')$ in
section 2, also applies to $\Sigma$ and $\Sigma'$ with even
boundaries, as above, since such $\Sigma$ are closed when identifying
coefficients mod $2$ \ie working over $Z_2$. The potentially confusing
possibility that intersection points may migrate out of $\Sigma$
through the boundary is actually harmless. This is because the fact
that the boundary is twice a cycle implies such a migration occurs in
pairs thereby preserving $J(\Sigma,\Sigma')$ mod $2$. Such $\Sigma$
with even (non-vanishing) boundary are not oriented and this is
likewise no problem as $1$ and $-1$ agree, mod $2$. 

It follows from all that we have said that the two quantities $w$ and $w^{(2)}$
are both elements of the same cohomology group
$H^2(\M_4,Z_2)$. Furthermore, when $\Sigma$ and $\Sigma'$ are closed,
the two definitions of their intersection number $I(\Sigma,\Sigma')$ and
$J(\Sigma,\Sigma')$ differ by an even integer. Consequently, using the
result of section 5, 
$$
w(\Sigma)=J(\Sigma,\Sigma),\quad \hbox{mod}\,\, 2 \quad 
\hbox{if}\,\,\,\partial\Sigma
=0.
$$
Comparison with Wu's formula, (6.5), then suggests that $w$, as
defined in our discussion, and the Stiefel-Whitney class, $w^{(2)}$,
are the same. Indeed this is true, as there is a mathematical result
stating that $w^{(2)}$ does satisfy our definition (4.8), see appendix A,
[Lawson and Michelsohn 1989].

It is remarkable that relatively straightforward physical arguments
involving complex spinor wave functions and the quest for a
generalised version of Stokes' theorem have led inexorably and
naturally to the initially abstruse concept of $Z_2$ cohomology.

Finally let us note a nice statement of Poincar\'e duality, following 
from (2.5) and the universal coefficient theorem that equates the  homology and
cohomology under consideration [see Schwarz 94]:
$$
H^2(\M_4,Z_2)\equiv H_2(\M_4,Z_2).
$$
\bigskip
\ni{\bf 7. Classification of four-manifolds relevant to
electro-magnetic duality}
\medskip
Our main result, representing a generalised version of the Dirac quantisation
condition in the presence of complex spinor wave-functions, is equation (6.4).
It applies  whenever the two-chain $\Sigma$ is either closed (and so
orientable) or possesses even boundary (and so is not orientable). The
quantity $w(\Sigma)$ appearing is well-defined modulo $2$ and can be
identified with an element of the Stiefel-Whitney class in
$H^2(\M_4,Z_2)$. 

We shall see how this result leads to a natural classification of all
relevant four-manifolds into three distinct classes, that we shall
denote I, II and III. (By relevant we mean connected, oriented,
compact and smooth and hence satisfying Poincar\'e duality). 

The first category, type I manifolds, are those for which all two-chains
$\Sigma$ with even (or vanishing) boundary have even $w(\Sigma)$. Then
the quantisation condition (6.4) reads 
$$
\e^{\im{q_F\over\hbar}\int_{\Sigma}F}=
\e^{2\im{q_F\over\hbar}\oint_{\alpha}A}.
$$
If $\Sigma$ is a two-cycle \ie if $\alpha$ vanishes, this yields
$$
{q_F\over 2\pi\hbar}\int_{\Sigma} F\in\Zs
\quad\forall\,\Sigma\in H_2(\M,\Zs),
$$
which is the same as the quantisation condition (1.1) with $q_F$
replacing $q_B$. If this is the only constraint then it is consistent
for all fluxes to vanish and hence all field strengths and all gauge
potentials. Alternatively it is consistent for the charge $q_F$ to vanish
which is the same as saying that the four-manifold admits neutral spinors, or
in mathematical terminology, possesses a spin structure. Notice that in this 
case I, the intersection matrix defined in section 2 is even, \ie has even 
entries on the diagonal. Thus the vanishing of $w$ mod $2$ leads to
the same conclusion as the vanishing of the Stiefel-Whitney class, as
it should, according to the mathematical result quoted earlier.

If $w$ is not always even, there are two possibilities, namely that it is
odd on some cycles (type II) or that it is even on all cycles so that
the oddness is visible only on some two-chains with even
boundary. This is the type III possibility. 

Let us consider the type II possibility. We already noted that
$w(\Sigma)$, being a self-intersection number $I(\Sigma,\Sigma)$, vanished on
torsion cycles. Hence the relevant cycle $\Sigma$ must be free
and the quantisation condition (1.2) implies that the flux through
$\Sigma$ is fractional. In particular it cannot vanish.
Nor can the spinor charge $q_F$ and hence neutral spinors are not admitted.
Mathematically it is said that the four-manifold possesses a spin$_C$
structure but not a spin structure. Notice that in this case the
intersection matrix for free cycles is odd in the terminology of section 5.

For type III four-manifolds the intersection matrix for free
two-cycles is even, yet, nevertheless, the gauge potential cannot be
gauge equivalent to zero even though the fields strengths $F$ may
vanish. The proof was given in the previous section.

To summarise, all three types support charged spinors, that is, spin$_C$
structures, but only type I manifolds support neutral spinors, that is, spin
structures.

On the other hand, only four-manifolds of types I and III possess even
intersection matrices for their free two-cycles. It is this distinction 
that will affect the way electromagnetic duality is realised.

The question of a likely relation between the two electric charges $q_B$ and
$q_F$ carried by the scalar and spinor wave functions has been deliberately
left open until now. There are two reasonable choices; either $q_F$
equals $q_B$ or half that. Given the first choice it follows from the
quantisation conditions (1.1) and (1.2) that on type I and III four-manifolds
(for which $I(\Sigma,\Sigma)$ is always even) that the scalar and spinor wave
functions coexist on the same backgrounds of fluxes. But on type II
four-manifolds the required backgrounds are incompatible so that there is no
choice for which both wave functions coexist. However it seems that it is this
choice that accords best with electromagnetic duality as we intend to explain
[M Alvarez and D Olive to appear]. With the second choice, in which $q_F$
equals one half $q_B$, the backgrounds that allow both wave functions are
precisely those required by the spinor wave function.

\bigskip
\ni{\bf 8. Conclusions and discussion}
\medskip
We believe the results clarify, as intended, the close relationship
between several notions concerning four-manifolds, the idea of spin  
structure, the Dirac quantisation condition for fluxes, the 
Atiyah-Singer index theorem for the $U(1)$ Dirac operator on 
the manifold and electromagnetic duality, not to mention the role of $Z_2$
cohomology. Evidently even such a simple gauge theory has a remarkably rich
structure.

However the discussion was limited to a certain class of four-manifold 
and there are reasons to think that the relevant class could be larger. 
Thus there are several ways in which it would be interesting and 
relevant to extend the results.

All the four-manifolds considered admit Riemannian (positive definite)
metrics but some admit Minkowski metrics and it would be worth 
knowing that the results extended to the corresponding Minkowski
spinors. This is very likely as most of the argument is topological
and so independent of metric. Those manifolds admitting Minkowski
metrics must have vanishing Euler number, where
$\chi(\M_4)=2-2b_1+b_2$, using Poincar\'e duality. Hence the first
Betti number has to be two or more if there are nontrivial fluxes and
so  such manifolds cannot be simply connected. 

A more urgent question concerns the extension to non-compact manifolds
as these are more interesting physically. In fact there is a modified
version of Poincar\'e duality that holds in these cases and this could
be relevant. See \eg [Schwarz 94]. Recall that the more physical
example of electromagnetic  duality [Montonen and Olive 1977] involves
a supersymmetric $SU(2)$ gauge theory in $R^{3,1}$ which is certainly
non-compact. 

Although electromagnetic duality of Maxwell fields is special to space-times of
four dimensions the Poincar\'e duality of the underlying space-time that seems
to be an essential prerequisite is not special to four dimensions. Hence
another interesting question concerns possible generalisations of our work 
to higher dimensions than four. This would have to involve higher 
order gauge potentials and wave functions for $p-$branes 
(whatever these wave functions are). Thus, on a (Poincar\'e dual and
connected) manifold $\M_{4k}$, of dimension $4k$, $2k$ field strengths 
could play a special role, as in the $k=1$ case investigated in this paper.
The $(2k-1)$-form gauge-potentials $A$ for which $F=dA$ could couple
to ``the wave function'' of a $(2k-2)-$brane. One might anticipate that
at least one important obstruction to the existence of wave functions
for ``spinning'' $(2k-2)-$branes would be supplied by an element
$w^{2k}\in H^{2k}(\M_{4k},Z_2)$, which is presumably the appropriate
Stiefel-Whitney class. If so, as $\delta w^{2k}=0$, mod $2$, either
$\delta w^{2k}=0$ or $\delta w^{2k}=2\lambda$ where $\lambda$ is a
$(2k+1)$-torsion-cocycle. Thus we have a lift either to
$H^{2k}(\M_{4k},\Zs)$ or $T^{2k+1}(\M_{4k},\Zs)$. But by equations
(2.4) and (2.5), 
$$
T^{2k+1}(\M_{4k},\Zs)=T_{2k}(\M_{4k},\Zs)=T^{2k}(\M_{4k},\Zs).
$$
So the same miracle that we saw for $k=1$ in section 4 repeats itself.
So maybe the existence of spinning $(2k-2)$-branes on general
(Poincar\'e dual) backgrounds actually requires the coupling to the
$2k-$form field strength originally noted by Kalb and Ramond [1974] in
the case of strings. 

In general,  these, and other, extensions of the sort of analysis in
this paper are likely to be important in a comprehensive theory of strings 
and branes moving in non-trivial backgrounds.

\bigskip 
{\bf Acknowledgments}
\medskip

We wish to thank the following for discussions; Tobias Ekholm,
Gary Gibbons, Stephen Howes, Robion Kirby, Nadim Mahassen and 
Albert Schwarz. We are grateful to the Mittag-Leffler Institute
for hospitality and to EPSRC and to TMR grant FMRX-CT96-0012 
for assistance.
\bigskip
\noindent {\bf References}
\medskip
\ni O Alvarez; \lq\lq Topological quantization and cohomology",
 {\it Comm Math Phys} {\bf 100} (1985) 279-309.

\ni E Corrigan and DI Olive; \lq\lq Colour and magnetic monopoles",
 {\it Nucl Phys} {\bf B110} (1976) 237.

\ni PAM Dirac; \lq\lq Quantised singularities in the electromagnetic field", 
{\it Proc Roy Soc} {\bf A33} (1931) 60-72.

\ni R Geroch 1968; \lq\lq Spinor structures of space-times in general
relativity I",{\it Journ Math Phys} {\bf 9} (1968) 1739-1744.

\ni R Geroch 1970; \lq\lq Spinor structures of space-times in general
relativity II",{\it Journ Math Phys} {\bf 11} (1970) 343-348.

 \ni P Goddard and DI Olive; \lq\lq Algebras, 
lattices and strings" in \lq\lq Vertex operators in mathematics and physics", 
MSRI publications {\bf 3} (Springer 1984) 51-96.

\ni N Habegger; \lq\lq Une variet\'e de dimension $4$ avec forme
 d'intersection paire et signature-8" {\it Commun Math Helv}
  {\bf 67} (1982) 22-24.
 
\ni SW Hawking and CN Pope; \lq\lq Generalised spinor structures 
in quantum gravity", {\it Phys Lett} {\bf B73} (1978) 42-44.

\ni M Kalb and P Ramond; \lq\lq Classical direct interstring action",
 {\it Phys Rev} {\bf D9} (1974), 2273-2284

\ni RC Kirby; \lq\lq The topology of 4-manifolds", (Springer Lecture Notes
 in Mathematics 1374, 1989).

\ni HB Lawson and M-L Michelsohn; \lq\lq Spin Geometry", 
{\it Princeton Mathematical Series} {\bf 38}, (Princeton 1989)

\ni J Milnor and D Husemoller; \lq\lq Symmetric bilinear forms"  
(Springer 1973)

\ni C Montonen and DI Olive; \lq\lq Magnetic monopoles as gauge fields?", 
{\it Phys Lett} {\bf 72B} (1977) 117-120.

\ni A Schwarz; \lq\lq Topology for Physicists", Springer 1994.

\ni E Verlinde; \lq\lq Global aspects of electric-magnetic duality",
 {\it Nucl Phys} {\bf B455} (1995) 211-228.

\ni E Witten; \lq\lq On S duality in abelian gauge theory", 
{\it Selecta Math
(NS)} {\bf 1} (1995) 383-410, hep-th/9505186.

\ni TT Wu and CN Yang; \lq\lq Concept of non-integrable phase factors 
and global formulation of gauge fields", {\it Phys Rev} {\bf D12} 
(1975) 3845-3857.
\bye